\documentclass[12pt,letterpaper]{article} 

\usepackage[utf8]{inputenc}
\usepackage[T1]{fontenc}
\usepackage{times} 
\usepackage{amsmath,amsfonts,amssymb}
\usepackage{graphicx}
\usepackage{setspace}
\usepackage[left=0.875in, right=0.875in, top=1.0in, bottom=1.25in]{geometry}

\usepackage{titlesec}
\titleformat{\section}{\normalfont\large\bfseries\uppercase}{\thesection.}{0.5em}{}
\titleformat{\subsection}{\normalfont\large\bfseries}{\thesubsection}{0.5em}{} %

\newcommand{\keywords}[1]{\noindent\textbf{Keywords:} #1}
 
\usepackage{amsmath,amsfonts,amssymb}
\usepackage{graphicx}
\usepackage[colorlinks=true, allcolors=blue]{hyperref}
\usepackage{authblk} 

\usepackage{fancyhdr}
\date{\today}        
\date{Octobre 2026}  

\title{FROM EARTH’S HARSHEST LANDSCAPES TO LUNAR HORIZONS \\ 
\large \textbf{Engineering NEA Telescopes for Extreme Environments}}

\author[a,b,c]{M. Hadjara}
\author[d]{A. Pomazan}
\author[a]{H. Ji}
\author[a]{Q. Dong}
\author[d]{J. Zheng}
\author[d]{M. Yindun}
\author[e]{L.Cament}
\author[e]{M. Adams}
\author[c]{M. Diaz}
\author[c]{R. Mendez}
\author[g,h]{B. Foing}
\author[i]{M. Meftah}
\author[j]{Z. Ping}
\author[k]{R.Ikhlef}
\author[k]{D.Baba Aissa}
\affil[a]{Nanjing Institute of Astronomical Optics \& Technology, National Astronomical Observatories, Chinese Academy of Sciences, Nanjing 210042, China}
\affil[b]{Chinese High Angular Resolution Southern Astronomical Laboratory (CHARSAL), Dept. of Astronomy,  Camino El Observatorio \#1515, Las Condes, Santiago, Chile. Postal Code: 7591245. Casilla 36-D, Santiago - Chile}
\affil[c]{Space and Planetary, Exploration Laboratory (SPEL), Dept. of Electrical Engineering, FCFM, Universidad de Chile, Av. Tupper 2007, Santiago, Chile. (E-mail: mhadjara@uchile.cl)}
\affil[d]{Shanghai Astronomical Observatory, Chinese Academy of Sciences, Shanghai, China}
\affil[e]{Dept. of Electrical Engineering, FCFM, Universidad de Chile, Av. Tupper 2007, Santiago, Chile}
\affil[f]{Dept. of Astronomy,  Camino El Observatorio \#1515, Las Condes, Santiago, Chile. Postal Code: 7591245. Casilla 36-D, Santiago - Chile}
\affil[g]{ILEWG (International Lunar Exploration Working Group), Vrije Universiteit Amsterdam, The Netherlands}
\affil[h]{ERA chair for space photonics , Latvia U , Riga and Leiden Observatory}
\affil[i]{LATMOS (Laboratoire Atmosphères, Milieux, Observations Spatiales), Université Paris-Saclay, France}
\affil[j]{Institute for Advanced Study, Shenzhen University, Shenzhen, China}
\affil[k]{Centre de Recherche en Astronomie, Astrophysique et Géophysique (CRAAG), Algiers, Algeria}

\begin{document} 
\maketitle

\thispagestyle{fancy} 
\pagestyle{fancy}     

\begin{abstract}
Near-Earth Asteroids (NEAs) represent both a scientific opportunity and a potential threat to human civilization. While current survey programs have successfully detected most large NEAs, a significant observational gap remains for objects smaller than approximately 140 m, which can still produce regional-scale devastation. Ground-based surveys are intrinsically limited by atmospheric conditions, weather, and solar avoidance constraints, which leave a significant fraction of near-Sun objects difficult to detect. In particular, Earth-based observatories remain largely blind to objects approaching from the Sun-ward direction, reducing available warning times.
Future lunar observatories offer a promising solution to this limitation. A telescope operating on the Moon could enable observations of regions inaccessible from Earth, provide enhanced sensitivity, and significantly improve early detection capabilities. However, the lunar environment imposes severe engineering constraints, including thermal cycling, dust contamination, radiation exposure, and autonomous operation requirements.
We propose the \textbf{AMAS (Atacama–Mongolia–Antarctica–Sahara) project}, an international initiative using Earth's most extreme environments as analogue sites for future lunar astronomical facilities. By deploying and testing specialized NEA survey telescopes in these environments, key technologies for future lunar observatories can be validated under realistic operational conditions. This paper presents the scientific rationale, engineering challenges, site selection strategy, and long-term roadmap toward a global Earth-Moon network for planetary defense.
\end{abstract}


\section{INTRODUCTION}
\label{sec:intro}  

Asteroid impacts have profoundly influenced the history of Earth, acting both as destructive agents and as fundamental drivers of planetary and biological evolution. Throughout the age of the Solar System, collisions with asteroids and other small bodies have repeatedly modified Earth's geology, atmosphere, climate, and biosphere. While impacts are often associated with catastrophic destruction, they have also played a crucial role in shaping the conditions that enabled the emergence and evolution of life \cite{Marchi2014}.

\subsection{Celestial Collisions: Driving Earth's Evolution and Life}
\label{sec:1.1}

One of the most significant epochs in Earth's early history was the Late Heavy Bombardment (LHB), approximately 4.1 billion years ago. This period is believed to have been triggered by orbital resonances between the giant planets, particularly Jupiter and Saturn, which destabilized large populations of asteroids and sent them toward the inner Solar System \cite{Gomes2005}. During this era, countless impacts reshaped the surfaces of the terrestrial planets. Asteroids may also have delivered a substantial fraction of Earth's water and volatile compounds, contributing to the conditions necessary for the emergence of life \cite{Morbidelli2000}.

Large impacts may have influenced biological evolution in ways that extend beyond simple destruction. Geological evidence associated with the S2 impact event, produced by an asteroid estimated to be between 30 and 60 km in diameter approximately 3.2 billion years ago, suggests that asteroid collisions could simultaneously generate severe environmental disturbances and chemically reactive environments favourable to prebiotic chemistry \cite{Lowe2025}. While such events may have locally sterilized portions of the Earth's surface through intense heating and ocean vaporization, they also released nutrients and energy sources capable of stimulating biological activity.

At the opposite extreme, asteroid impacts have repeatedly caused major biological crises. The most famous example is the Chicxulub impact approximately 66 million years ago. The impactor, estimated to have been 10--12 km in diameter, excavated a crater approximately 180 km wide on the present-day Yucat\'an Peninsula in Mexico \cite{Schulte2010}. The resulting global environmental catastrophe contributed to the extinction of nearly 75\% of Earth's species, including all non-avian dinosaurs.

\subsection{Near-Earth Asteroid Impacts on Earth: Shaping Human Life and Culture}
\label{sec:1.2}

Asteroids have also influenced human civilization, both directly and indirectly. Beyond their geological and climatic effects, celestial impacts have inspired myths, legends, religious interpretations, and cultural narratives throughout human history.

Although civilization has never experienced a Chicxulub-scale event, several recent impacts illustrate the continuing hazard posed by near-Earth asteroids (NEAs). The Tunguska event of 1908, likely produced by a 50--80 m asteroid or cometary fragment, generated an atmospheric explosion equivalent to approximately 15 megatons of TNT \cite{Boslough2008}. The blast flattened more than 2,000 km$^2$ of Siberian forest.

More recently, the Carancas meteorite impact in Peru in 2007 demonstrated that even relatively small rocky bodies can survive atmospheric entry and reach the ground \cite{Tancredi2009}. Although the object was only a few meters in diameter, it produced a crater approximately 13 m wide.

The Chelyabinsk meteor of 2013 provided another reminder of the threat posed by small near-Earth objects (NEOs). The object, approximately 20 m in diameter and with an estimated mass of about 10,000 tons, exploded at an altitude of about 30 km with an energy release of nearly 500 kilotons of TNT \cite{Brown2013, Popova2013}. The resulting shock wave injured more than 1,500 people and significantly damaged local infrastructure. This event also highlighted the limitations of current survey programs in monitoring objects approaching from Sunward directions.

Today, planetary defence has become a major international scientific and technological priority. Survey systems such as Pan-STARRS, Catalina Sky Survey, ATLAS, and the newly commissioned Vera C. Rubin Observatory continuously monitor the sky, detecting and characterizing NEAs and particularly  potentially hazardous asteroids (PHAs)\cite{Jedicke2016, Mainzer2020}. Nevertheless, a significant fraction of intermediate-to-small sized objects remains undiscovered  or lost.

The challenge facing planetary defence is therefore not merely the detection of hazardous objects, but their detection sufficiently early to enable and implement effective mitigation strategies \cite{Harris2015}.

\section{The NEA Detection Gap}
\label{sec:2}

Current statistics indicate that the NEAs population is increasingly well characterized at large asteroids, but remains highly incomplete at smaller scales. According to the Minor Planet Center (MPC) of the International Astronomical Union (IAU), more than 1.5 million small Solar System objects (SSOs) have been identified, including over 42,000 NEAs (Table~\ref{tab:1}). Among these, there are only a few hundreds objects with diameters more than 1 km, while several thousand are classified as PHAs \cite{MPC2026, Harris2015}.

\begin{table}[h]
\centering
\caption{Current population statistics of small SSOs (IAU Minor Planet Center).}
\begin{tabular}{lc}
\hline
Type & Number \\
\hline
Total number & $\sim$1,500,000 \\
Numbered minor planets & $>$850,000 \\
Unnumbered minor planets & $>$650,000 \\
Near-Earth asteroids (NEAs) & $>$42,000 \\
NEAs $>$ 1 km & $\sim$900 \\
Potentially hazardous asteroids (PHAs) & $>$2,500 \\
Comets & $>$4,600 \\
\hline
\end{tabular}
\label{tab:1}
\end{table}

\textit{PHAs} are defined as objects larger than approximately 140 m in diameter with a minimum orbital intersection distance (MOID) with Earth of less than 0.05 AU, as such objects in case of Earth impact will cause significant regional destruction.

\begin{figure}[h]
\centering
\includegraphics[width=0.9\linewidth]{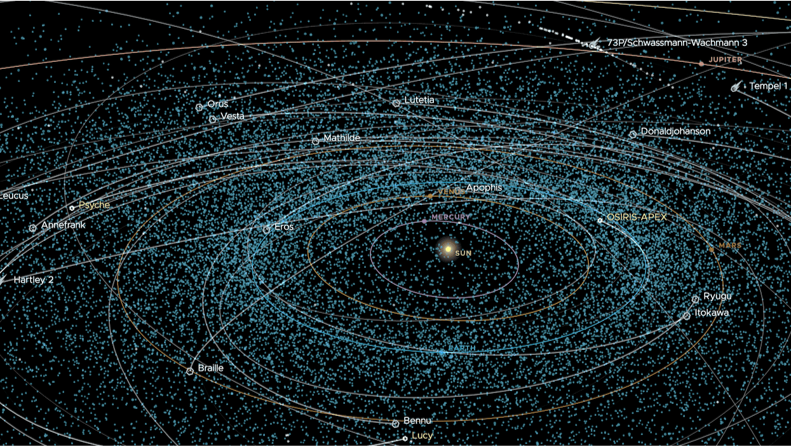}
\caption{Orbital distribution of some known near-Earth objects and selected spacecraft missions for asteroid and comet exploration.}
\label{fig:1}
\end{figure}

Despite significant progress in survey completeness, detection efficiency decreases rapidly for smaller objects, particularly below 300m \cite{Harris2015, Jedicke2016}. Figure \ref{fig:1} illustrates the orbital distribution of some known near-Earth objects and highlights several spacecraft missions dedicated to the exploration of asteroids and comets.

Three main impact-risk scenarios can be distinguished \cite{b612annualreport2020}:

\begin{itemize}
\item \textbf{Large NEAs ($>$ 1 km):} global catastrophe potential; more than 95\% already discovered and tracked; extremely rare impact frequency (order of $10^5$--$10^6$ years).
\item \textbf{Medium NEAs (140 m -- 1 km):} capable of regional to continental-scale devastation; partial census completeness; impact frequency on the order of $10^4$ years ($\backsim$40\% discovered/tracked).
\item \textbf{Small NEAs ($<$ 140 m):} city-scale destruction events (e.g., Tunguska, Chelyabinsk); majority remain undetected; characteristic impact time-scale of decades to centuries ($\backsim$10\% discovered/tracked).
\end{itemize}

This distribution defines the so-called \textbf{detection gap}, i.e., the incomplete observational census of sub-kilometre objects that still pose a significant risk to human society. While current surveys have identified approximately 95\% of kilometre-scale NEAs, the discovery completeness decreases rapidly for smaller, but still potentially hazardous objects, reaching only about 40\% for NEAs between 140~m and 1~km and about 10\% for objects between 45~m and 140~m, leaving approximately 90\% of the latter population still undetected.

Several observational limitations contribute to this gap, including atmospheric extinction, weather conditions, daylight constraints, limited sky coverage, and survey cadence limitations. These effects introduce strong observational biases that limit completeness, particularly for fast-moving objects and those with short visibility windows \cite{Jedicke2016}.

Among these limitations, the most critical is the Sun-ward blind zone. Ground-based observatories cannot observe regions close to the Sun due to solar avoidance constraints and increased background brightness. As a consequence, asteroids approaching from solar directions often remain undetected until shortly before close approach or impact \cite{Mainzer2020}.

Historically, several objects illustrate this limitation, for example 2008 TC$_3$ and 2019 OK, as well as multiple other NEAs detected only days before their closest approach to Earth.

Overcoming these constraints and blind regions remains one of the primary motivations for future space-based and lunar survey architectures, which aim to enable sustained high-cadence monitoring of near-Sun regions inaccessible to ground-based observatories.

\section{Why the Moon Changes the Game}
\label{sec:3}

A telescope operating on the lunar surface offers unique observational advantages \cite{Hadjara2026SPIE}, particularly for NEA detection and planetary defence. Unlike Earth-based facilities, a lunar observatory benefits from the absence of an atmosphere, reduced scattered light, and extended access to near-Sun regions that are challenging to observe from Earth. Indeed, it can provide enhanced coverage of sky regions close to the Sun, which are largely inaccessible from ground-based observatories due to solar glare and sky background brightness constraints \cite{Foing2023}.

\subsection{Extended Warning Time from Lunar Geometry}
\label{sec:3.1}

NEAs observation is critical not only for planetary defence, but also for scientific characterization and future resource utilization. Specialized survey systems are therefore required for both regular follow-up monitoring and time-crucial alert observations.

Given the average relative velocity of NEAs approaching Earth ($\approx 20~\mathrm{km,s^{-1}}$), the warning time could be extended over a few hours for cases involving the detection of imminent impactors.

Although this time interval may appear modest, it is scientifically and operationally significant for planetary defence scenarios, particularly for small impactors where early warning is essential to mitigate regional threats. Depending on the encounter geometry, lunar-based survey systems may provide up to 5--6 hours of additional warning time compared to Earth-based observatories. Importantly, this gain occurs during the critical final-approach phase, when impact probabilities are refined, impact corridors are constrained, and timely alerts and mitigation actions can be initiated.

\subsection{The Sun-ward Advantage: Breaking the Earth Blind Spot}
\label{sec:3.2}

One of the most fundamental limitations of ground-based asteroid surveys is the Sun-ward blind zone. Objects approaching from directions close to the Sun remain largely undetectable due to atmospheric scattering and increased sky background brightness, which reduce the sensitivity of ground-based telescopes in these regions.

A lunar-based observatory could notably this limitation. From the Moon, solar elongation constraints are relaxed, enabling earlier detection of incoming NEAs in geometries that are inaccessible from Earth.

Table \ref{tab:2} characterizes the first-detection cases for the exemplified asteroids in each group for the  thermal infra-red observations.

\begin{table}[h]
\centering
\caption{Estimated first detection distance and corresponding warning time for representative NEAs, assuming a relative velocity of $v \approx 20~\mathrm{km\,s^{-1}}$.}
\begin{tabular}{lcc}
\hline
Object size & First detectable distance & Extra lead time \\
\hline
30 m (city-scale impactor) & 0.05 AU ($\sim 7.5 \times 10^6$ km) & $\sim$4--5 days \\
140 m (potentially hazardous) & 0.10 AU ($\sim 1.5 \times 10^7$ km) & $\sim$9 days \\
1 km (global-scale impactor) & 0.50 AU ($\sim 7.5 \times 10^7$ km) & $\sim$6 weeks \\
\hline
\end{tabular}
\label{tab:2}
\end{table}

\noindent
These values are given for reference only and depend strongly on albedo, phase angle, instrument aperture, and sky background conditions. However, they illustrate the challenging detectability of small NEAs at large distances and low apparent brightness, highlighting the need for improved survey capabilities for objects approaching from poorly observed regions of the sky.

\subsection{Why Specialized NEA Telescopes Are Critical}
\label{sec:3.3}

Planetary defence requires dedicated observational facilities optimized for the early detection and follow-up of small or fast-moving objects. General-purpose observatories can complement these facilities by providing additional observations, increasing data coverage, and extending the orbital arcs required for accurate orbit determination.

The rationale for specialized NEA observatories can be summarized as follows:

\begin{itemize}
\item \textbf{Planetary defence:} Early detection, impact probability assessment, and object characterization enable mitigation strategies for potentially hazardous impactors, particularly small NEAs (10--100 m), which are often detected only after close approach.
\item \textbf{Dedicated survey capability:} Specialized NEA observatories provide optimized cadence, sensitivity, and scheduling flexibility to detect and track small and fast-moving objects that are challenging for general-purpose observatories.
\item \textbf{Scientific return:} NEA observations provide key constraints on Solar System evolution, composition, and dynamical history.
\item \textbf{Resource potential:} NEAs represent potential future sources for in-situ resource utilization and space economy development.
\end{itemize}

\section{Earth's Harshest Testbeds}
\label{sec:4}

Before deployment on the lunar surface, next-generation NEA survey systems must be validated in extreme terrestrial environments that provide operational analogues for autonomy, thermal stability, optical and mechanical performances, and long-duration reliability.

Desert sites such as the Atacama, Sahara, and Gobi offer exceptional atmospheric transparency and low humidity, but impose strong thermal gradients and dust exposure, requiring sealed and thermally stable instrument designs.

High-altitude observatories in Mauna Kea, the Tibetan Plateau, and the Andes provide reduced atmospheric absorption and excellent seeing conditions, while introducing constraints in accessibility and environmental harshness that drive remote operation capabilities.

Polar facilities, particularly in Antarctica (Dome A, Dome C, South Pole), represent the closest analogue to lunar conditions, combining long-duration observing conditions and extended darkness with extreme thermal conditions logistical limitations that necessitate full system autonomy.

Together, these environments form a progressive validation pathway toward lunar deployment, each addressing key constraints expected on the Moon, including thermal cycling, autonomy, environmental sealing, and operational robustness. In summary, each class of site maps to a specific set of technological challenges:

\begin{itemize}
\item \textbf{Desert sites:} high optical quality balanced against dust exposure and pollution, and strong thermal gradients,
\item \textbf{High-altitude sites:} improved atmospheric performance at the expense of accessibility and operational constraints,
\item \textbf{Polar sites:} long-duration observing conditions and darkness under extreme isolation, temperature and logistical limitations.
\end{itemize}

The Moon offers a unique platform for NEA detection by extending observational geometry beyond Earth limitations, increasing early warning time by several hours, and enabling access to Sun-ward directions that are inaccessible from the ground. 

This progression naturally leads from Earth-based analogues to lunar horizons, outlining the pathway toward the next generation of planetary defence infrastructure.

\subsection{The AMAS Concept}
\label{sec:4.1}

To prepare future lunar observatories, technologies must first be validated under harsh terrestrial conditions. The \textbf{AMAS} (\textbf{A}tacama, \textbf{M}ongolia, \textbf{A}ntarctica, and \textbf{S}ahara) project proposes a distributed testing framework based on four extreme environments:

\begin{itemize}
\item Atacama Desert (Chile)
\item Inner Mongolia (China)
\item Antarctica
\item Sahara Desert (North Africa)
\end{itemize}

These four sites were selected to reproduce, on Earth, a broad spectrum of lunar-relevant environmental conditions. Each location contributes uniquely to the validation of future autonomous astronomical infrastructures under extreme conditions.

The sites provide complementary test conditions spanning:

\begin{itemize}
\item Extreme dryness
\item Large temperature gradients
\item Dust contamination
\item Strong winds
\item Isolation
\item Autonomous operation requirements
\end{itemize}

Together, they form a natural terrestrial laboratory for future lunar astronomy technologies, enabling progressive testing of systems in conditions essentially close to those expected on the Moon.

\begin{figure}[ht]
\centering
 \includegraphics[width=0.9\linewidth]{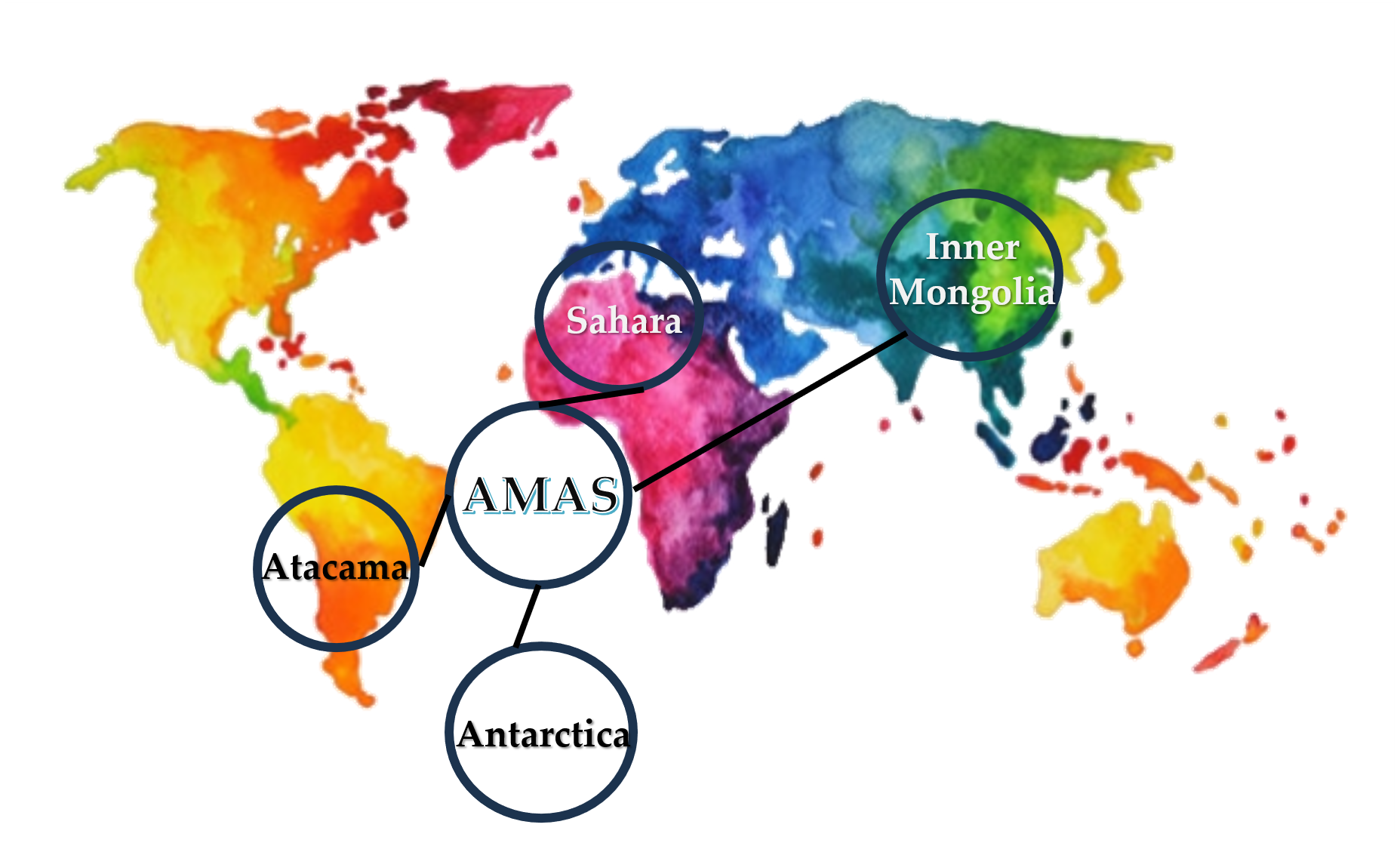}
\caption{Global distribution of the AMAS (Atacama–Mongolia–Antarctica–Sahara) test sites forming a distributed terrestrial analog network for validating technologies for future lunar observatories.}
\label{fig:2}
\end{figure}

Figure~\ref{fig:2} illustrates the global distribution of the AMAS test sites and highlights their complementary geographical and environmental coverage.

These sites collectively cover the full spectrum of challenges expected for lunar astronomy, including optical stability, thermal extremes, dust contamination, and long-duration autonomous operations. This distributed approach allows progressive validation of both hardware and software systems under increasingly realistic conditions prior to deployment on the Moon.

The Atacama Desert is among the driest places on Earth and hosts some of the world's premier astronomical facilities. It is characterized by excellent seeing conditions, low atmospheric water vapour content, high optical transparency, and the presence of existing observatory infrastructure. As a result, the site provides an ideal environment for validating optical systems as well as robotic operations.

The Inner Mongolia region offers cold winters, dust-rich conditions, and strong seasonal variability. These characteristics make it particularly valuable for testing actuators, enclosures, and contamination-control strategies under realistic environmental stress.

Antarctica provides perhaps the closest terrestrial analogue to certain lunar environmental conditions. It is characterized by extremely low temperatures, long polar nights, and exceptional atmospheric stability. In addition, the logistical constraints and need for high autonomy closely resemble those expected for future lunar facilities.

The Sahara Desert offers extreme heat, abrasive dust, and large thermal gradients. These conditions are particularly relevant for evaluating thermal control systems, structural resilience, and protective coatings under harsh operational environments.

Table~\ref{tab:3} summarizes the main environmental conditions, the corresponding on-site experimental activities, and their direct relevance to future lunar observatory deployment.

\begin{table}[ht]
\centering
\renewcommand{\arraystretch}{1.3}
\begin{tabular}{|p{3.cm}|p{4.cm}|p{4.cm}|p{4.cm}|}
\hline
\textbf{Site} & \textbf{Main testable constraints} & \textbf{Performed tests} & \textbf{Terrestrial analogue methods for lunar applications} \\
\hline

Atacama (Chile) &
Excellent seeing, dry air, clear nights &
Twilight detection (low solar elongation), thermal IR (3--5 µm), precise astrometry &
NEA pipelines (low S/N), non-sidereal tracking, optical stability \\
\hline

Inner Mongolia desert (China) &
Dark skies, large temperature swings, dust &
Dust resistance, dry lubricants, mechanical joints under cold/wind effects &
Lunar regolith mitigation, mechanical reliability \\
\hline

Antarctica (Dome C / A) &
Pristine skies, Ultra-low turbulence, extreme low temperatures, long polar nights, dry air &
Continuous observation, full autonomy, low energy consumption &
Autonomous operations, latency handling, survival through long lunar nights \\
\hline

Sahara &
Clear skies, extreme heat, abrasive dust &
Thermal stress, optical protection, solar power + batteries &
Thermal robustness, sealing, energy and dust management \\
\hline

\end{tabular}
\caption{Comparison of AMAS terrestrial analogue sites and their relevance for validating technologies for future lunar observatories.}
\label{tab:3}
\end{table}

\section{ENGINEERING REQUIREMENTS FOR FUTURE LUNAR NEA TELESCOPES}
\label{sec:5}

Future lunar-based telescopes dedicated to NEAs detection impose a set of stringent engineering constraints driven by the need for long-term autonomy, environmental resilience, and high-cadence wide-field observations. The following subsections summarize the main subsystems and their associated requirements.

\subsection{Optical System}
\label{sec:5.1}

A baseline concept consists of a compact wide-field telescope optimized for NEA survey operations, combining wide sky coverage, deep sensitivity to detect small and faint objects, and high angular resolution for precise astrometric measurements and impact probability assessment. Typical designs consider apertures ranging between 0.5 and 1 meter, combined with a field of view ideally around 5 degrees in order to balance sky coverage, image resolution and detectability of faint objects.

Such systems are primarily optimized for rapid sky surveys, enabling efficient detection of fast-moving objects rather than detailed physical characterization. Optical stability, lightweight structures, and minimal alignment drift are therefore key design drivers under lunar thermal cycling conditions. Sun protection baffles are also required to ensure safe operation near the solar limb and to minimize stray light contamination during low solar elongation observations.

\subsection{Detectors}
\label{sec:5.2}

Detector systems must ensure high sensitivity and robustness under low-flux conditions. Key requirements include high quantum efficiency, low readout noise, and fast cadence imaging capabilities to enable continuous monitoring of transient and fast-moving NEAs.

Both visible and near-infrared architectures are considered. Infrared observations in the 3--5 µm range are particularly attractive for detecting low-albedo asteroids, which are poorly constrained in visible wavelengths due to their reduced reflectivity. Radiation hardness and thermal stability of detector arrays are also critical for long-duration lunar deployment.

\subsection{Mechanical Design}
\label{sec:5.3}

Mechanical subsystems must be designed to operate reliably under abrasive dust exposure, vacuum conditions, and repeated thermal expansion cycles. Long-term autonomous operation without human intervention further increases the importance of structural robustness and redundancy.

The system must include dust-resistant motorized protective hoods to prevent particulate contamination during both idle and active operation phases. All optical and mechanical interfaces must rely on watertight seals and abrasion-resistant coatings to ensure long-term durability under harsh lunar-like conditions. Dry or cryogenic lubricants are preferred for all moving components in order to mitigate vacuum-induced degradation. In addition, optical fouling sensors are required to monitor dust accumulation on critical surfaces in real time and trigger corrective actions.

\subsection{Thermal Control}
\label{sec:5.4}

Thermal stability is a critical requirement for both optical alignment and detector performance. Potential solutions include multi-layer insulation, passive radiators optimized for deep-space heat rejection, and controlled heating elements to stabilize sensitive components during extreme environmental cycles.

Phase-change materials may also be considered to buffer temperature fluctuations over long timescales. The combination of passive and semi-active thermal control strategies is expected to provide the best compromise between robustness, power efficiency, and long-term stability.

\subsection{Software and Autonomy}
\label{sec:5.5}

Software autonomy constitutes a central pillar of system design and operational resilience. The telescope control system must support autonomous non-sidereal tracking over a wide range of stable angular rates, up to approximately 1.5 arcsec/sec, enabling observations of NEAs approaching from different directions with varying apparent motions. In parallel, the astrometric reduction software must perform precise centroiding measurements of trailed stars in the field of view (FOV) to ensure accurate orbit determination.

Robust self-recovery mechanisms, duplication of highly important nodes and subsystems are required, including cold reboot procedures, safe-mode transitions, and comprehensive anomaly logging to ensure continuous operation in case of system failures. Communication constraints are simulated through Moon--Earth latency models on the order of 2--3 seconds, requiring fully autonomous onboard decision-making.

Beyond these core functions, the system is expected to operate a fully autonomous processing pipeline, from raw observations to MPC report submission, including real-time generation of alerts for transient or potential impactors. This includes onboard detection, validation, and prioritization of candidates without ground intervention.

In addition, software autonomy extends to operational management, including observation scheduling, real-time data reduction, and adaptive decision-making module based on system health and environmental constraints.

Energy-aware and battery-conscious operation is also an integral part of the software layer. Task prioritization and computational load management must dynamically adapt to available power resources, particularly during low-insolation periods or extended eclipse-like conditions, ensuring continuous scientific operation under constrained energy budgets.

Consequently, artificial intelligence is absolutely essential for managing all the software- and autonomy-related aspects mentioned above.

\section{FROM FUTURE GENERATION OF SPACE TELESCOPES TO FUTURE LUNAR OBSERVATORIES}
\label{sec:6}

An alternative strategy for accelerating the development of lunar astronomy consists in leveraging the technological heritage of previous space observatories. However, directly recycling existing telescopes such as the Hubble Space Telescope for installation on the lunar surface is neither technically realistic nor economically attractive. Such instruments were designed for operation in Earth orbit and would require extensive modifications to withstand lunar dust, surface gravity, thermal cycling, and long-term autonomous operation.

A more practical approach consists of reusing proven technologies, subsystems, and engineering solutions developed for previous space missions while designing a new generation of instruments specifically optimized for the lunar environment. This strategy allows the benefits of decades of technological development to be retained while avoiding the limitations associated with adapting legacy systems beyond their original design specifications.

Potential advantages include:

\begin{itemize}
\item Leveraging heritage technologies and operational experience from previous space telescopes, thereby reducing development costs, technical risks, and deployment schedules;
\item Developing dedicated lunar-compatible technologies to address environmental and operational constraints, enhancing system robustness, autonomy, and long-term reliability.
\end{itemize}

\begin{figure}[ht]
\centering
\includegraphics[width=0.9\linewidth]{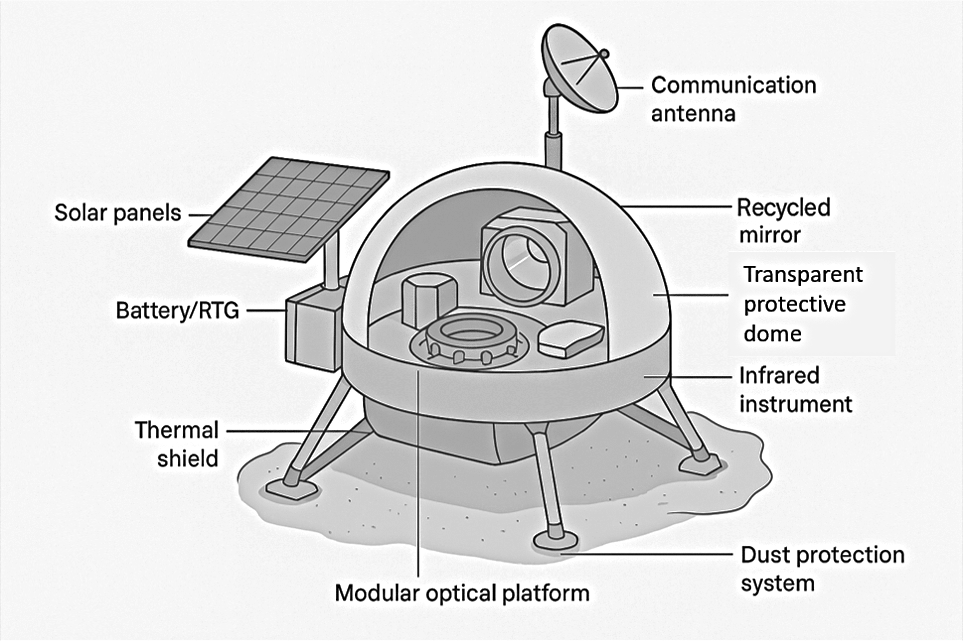}
\caption{The proposed architecture assumes a protected and autonomous observatory environment in which optical instruments operate within a controlled enclosure. The use of transparent protective structures, re-purposed mirrors (or segmented mirrors) from previous space missions introduces additional engineering challenges, including optical alignment, wavefront control, and thermal stability. These challenges represent key technology developments required for future lunar observatories.}
\label{fig:3}
\end{figure}

Within this framework, a particularly promising concept is the development of a standardized lunar observatory platform, conceived as a sealed and autonomous ecosystem that protects scientific instruments from lunar dust, vacuum exposure, and extreme thermal variations while providing long-term operational support under lunar conditions. Such a platform would provide the common infrastructure required by future telescopes, including thermal protection, power generation and storage, communication systems, structural support, stabilization capabilities, and autonomous control functions. Figure~\ref{fig:3} presents the concept of a lunar-based observatory platform and illustrates a conceptual architecture for a new generation of modular autonomous lunar telescopes based on a common infrastructure platform supporting interchangeable scientific payloads.

Rather than designing each observatory as a unique system, future instruments could be integrated onto a common lunar platform capable of accommodating different payloads according to mission objectives. This modular philosophy would significantly reduce development costs while enabling rapid deployment of successive generations of telescopes.

Such an approach would effectively establish a standardized "lunar observatory bus (or platform)" analogous to the spacecraft bus concept widely used in satellite engineering. Future optical, infrared, radio, or multi-wavelength instruments could then be developed independently and integrated onto this common infrastructure. A generic modular lunar observatory architecture based on a sealed integrated platform could therefore consist of:

\begin{itemize}
\item Telescope or scientific payload module;
\item Power generation and energy storage module;
\item Communication and data relay module;
\item Autonomous control and artificial intelligence module;
\item Thermal regulation module;
\item Dust mitigation and environmental protection module;
\item Mobility or deployment module for future expandable systems.
\end{itemize}

The combination of a stable lunar platform, sustained autonomous operations, and dedicated surveys of low-elongation Sunward regions could therefore open a new observational domain largely inaccessible to current ground-based facilities.

\section{A GLOBAL EARTH--MOON PLANETARY DEFENCE NETWORK}
\label{sec:7}

The ultimate objective extends far beyond the deployment of individual telescopes. The long-term vision is the establishment of an integrated Earth--Moon planetary defence network capable of continuously monitoring the near-Earth environment and providing early warning of potentially hazardous objects. Such a system should integrate Earth-based observatories, lunar facilities, autonomous processing centres, and AI-driven processing algorithms into a unified planetary defence ecosystem.

This approach is based on the principle that no single observatory, whether Earth-based or lunar, can provide complete coverage of the NEA population. Instead, a distributed network exploiting the complementary advantages of multiple observing locations offers the highest probability of detecting hazardous objects as early as possible and thus limiting their potential impact.

\subsection{Earth Segment}
\label{sec:7.1}

The terrestrial segment would remain the foundation of planetary defence activities. Existing and future survey facilities provide large sky coverage, mature operational infrastructures, and extensive experience in automated asteroid detection.

Representative facilities include:

\begin{itemize}
\item Rubin Observatory;
\item Pan-STARRS;
\item ATLAS;
\item AMAS test-bed observatories.
\end{itemize}

Ground-based observatories are particularly effective for large-scale sky surveys, rapid follow-up observations, orbit refinement, and long-term monitoring of known objects. The AMAS network described previously would additionally serve as a technology validation platform for future lunar systems while simultaneously contributing to planetary defence observations.

The geographical distribution of these facilities enables continuous monitoring over multiple longitudes and environmental conditions, reducing observational gaps caused by weather, daylight, and seasonal effects.

\subsection{Lunar Segment}
\label{sec:7.2}

The lunar segment represents the next development step in planetary defence infrastructure. By operating above Earth's atmosphere and from a different observational geometry, lunar observatories can access regions of the sky that are difficult or impossible to observe from terrestrial facilities.

Potential assets include:

\begin{itemize}
\item Near-side survey telescopes;
\item Polar observatories;
\item Infrared monitoring stations.
\end{itemize}

Near-side survey telescopes could conduct continuous searches for NEOs at low solar elongations. Lunar polar observatories would benefit from unique illumination conditions and potentially stable thermal environments, particularly near permanently shadowed regions. Dedicated infrared stations would provide complementary measurements of asteroid thermal emission, improving estimates of size, albedo, and impact energy.

Particularly important is the ability of lunar observatories to detect objects belonging to populations such as Atiras and other objects that spend much of their orbit close to the Sun from the terrestrial point of view. Earlier detection of such bodies could increase warning times by months or even years in some scenarios.

In the longer term, multiple lunar observatories could be deployed as a distributed network across different lunar latitudes and longitudes, providing redundancy, improved sky coverage, and enhanced resilience against individual system failures.

\subsection{Data Segment}
\label{sec:7.3}

The effectiveness of the Earth and lunar segments ultimately depends on the performance of the associated data infrastructure. Future planetary defence systems will increasingly rely on automated processing pipelines capable of handling massive data volumes with minimal human intervention.

Key components include:

\begin{itemize}
\item Autonomous detection pipelines;
\item Orbit determination systems;
\item AI-assisted risk assessment.
\end{itemize}

Autonomous detection pipelines would process observations in near real time, identifying moving objects and filtering false positives. Orbit determination systems would continuously update orbital solutions as new observations become available, allowing rapid refinement of impact probabilities.

Artificial intelligence techniques could assist in candidate classification, anomaly detection, observation scheduling, and impact-risk evaluation. Machine learning algorithms may also help prioritize follow-up observations by identifying objects with unusual dynamical properties or elevated hazard potential.

The integration of terrestrial and lunar observations within a common data architecture would provide a substantial increase in orbital accuracy and detection completeness compared with isolated observing systems.

Ultimately, the network could evolve toward a fully autonomous planetary defence ecosystem capable of detecting, characterizing, tracking, and assessing potentially hazardous objects with minimal delay between observation and issuing the alerts.

\begin{figure}[ht]
\centering
\includegraphics[width=0.95\linewidth]{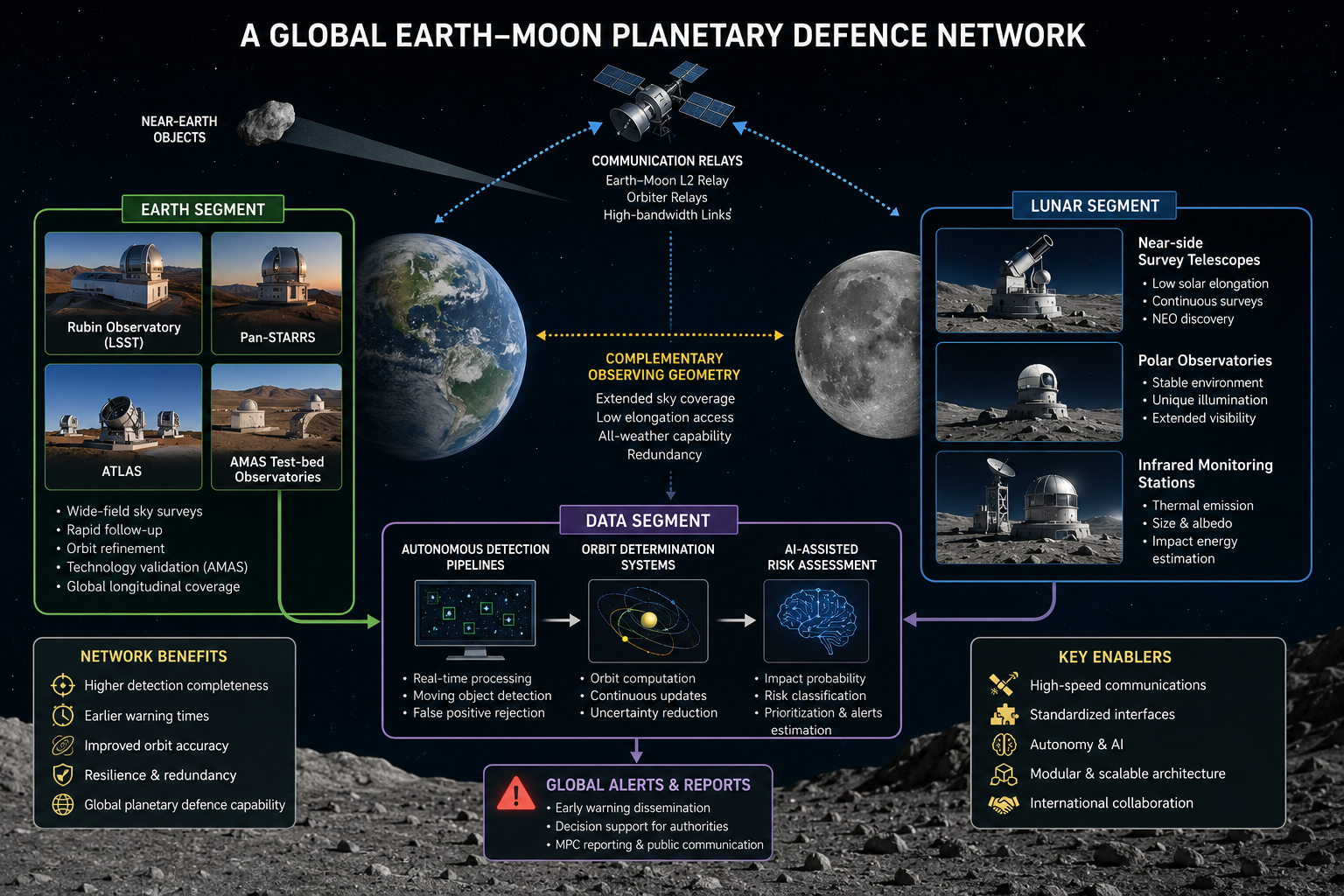}
\caption{Conceptual architecture of a future Earth--Moon planetary defence network. Ground-based observatories, lunar survey telescopes, infrared monitoring stations, autonomous processing pipelines, and AI-assisted risk assessment systems operate as a coordinated global infrastructure for Near-Earth Object detection and early warning.}
\label{fig:4}
\end{figure}

Figure~\ref{fig:4} illustrates a conceptual architecture of a future Earth--Moon planetary defence network combining terrestrial observatories, lunar facilities, communication relays, autonomous processing centers, and decision-support systems.

Such a multi-layer architecture has the potential to significantly improve detection completeness, orbital accuracy, and increase the warning times for hazardous objects. By combining the strengths of terrestrial and lunar astronomy, it may constitute a major step toward a truly global planetary defence facility for the twenty-first century.

\section{CONCLUSIONS}
\label{sec:Conclu}

NEAs remain both scientifically valuable targets and potential hazards for humanity. Over the past decades, major advances in survey capabilities, detector technologies, and automated data processing have dramatically increased the discovery rate of NEAs, and particularly potentially hazardous objects. Nevertheless, important observational gaps persist for small and medium-sized asteroids approaching from low solar elongations or from regions of the sky that remain difficult to observe from Earth.

The growing awareness of planetary defence challenges highlights the need for complementary observational strategies capable of extending current detection capabilities. In this context, lunar observatories offer a particularly promising opportunity. Their unique observing geometry, absence of atmosphere, and potential for continuous autonomous operation could significantly improve the detection of objects approaching from the Sunward direction, including Atira-like asteroids that are currently underrepresented in existing surveys. Earlier detection of such objects could translate into substantially longer warning times and more effective mitigation strategies for the potential impact scenarios.

This paper has outlined a progressive roadmap toward such a capability. Rather than directly pursuing complex and costly lunar deployments, the proposed AMAS concept provides an intermediate and pragmatic step based on terrestrial analogues. By exploiting the complementary environments of the Atacama Desert, Inner Mongolia province of China, Antarctica, and the Sahara desert, critical technologies can be tested and validated under conditions that reproduce many of the operational challenges expected on the Moon surface. These include thermal extremes, dust contamination, long-duration autonomy, energy management, and remote control operations.

The AMAS framework also offers benefits beyond technology validation. As a distributed international network, it provides opportunities for scientific observations, operational training, international collaboration, and the development of autonomous observing systems that can later be transferred to lunar applications. The project therefore serves both as a planetary-defence demonstrator and as a precursor to future lunar astronomical infrastructures.

A key conclusion emerging from this study is that future lunar observatories should not necessarily be conceived as isolated flagship missions. Instead, a modular architecture based on standardized lunar platforms may provide a more sustainable and scalable approach. Such platforms could support successive generations of optical, infrared, and multi-wavelength instruments while benefiting from common infrastructure for power, communications, thermal control, and autonomous operations. This philosophy could substantially reduce development costs while accelerating the deployment of a permanent astronomical presence on the lunar surface.

Ultimately, the long-term vision extends toward the creation of a fully integrated Earth--Moon planetary defence network. By combining terrestrial survey facilities, AMAS test-bed observatories, future lunar telescopes, autonomous processing pipelines, and artificial intelligence-assisted decision systems, such a network could provide unprecedented detection completeness, orbit determination accuracy, and early-warning capabilities. Beyond planetary defence, this infrastructure would also open new scientific opportunities in time-domain astronomy, infrared observations, and the study of previously inaccessible populations of Solar System objects.

As humanity enters a new era of lunar exploration and permanent cislunar (between Moon and Earth) activity, the Moon is increasingly emerging not only as a destination, but also as a strategic astronomical platform. The development of autonomous lunar NEA observatories, supported by terrestrial precursor networks such as AMAS, may therefore represent one of the most practical and scientifically rewarding pathways toward the next generation of planetary-defence and astronomical infrastructures.

\bibliography{report} 

@article{Gomes2005,
  author = {Gomes, R. and Levison, H. F. and Tsiganis, K. and Morbidelli, A.},
  title = {Origin of the cataclysmic Late Heavy Bombardment period of the terrestrial planets},
  journal = {Nature},
  volume = {435},
  number = {7041},
  pages = {466--469},
  year = {2005},
  doi = {10.1038/nature03676}
}

@article{Marchi2014,
  author = {Marchi, S. and Bottke, W. F. and Elkins-Tanton, L. T. and Bierhaus, M. and Wuennemann, K. and Morbidelli, A. and Chapman, C. R.},
  title = {Widespread mixing and burial of Earth's Hadean crust by asteroid impacts},
  journal = {Nature},
  volume = {511},
  number = {7511},
  pages = {578--582},
  year = {2014},
  doi = {10.1038/nature13539}
}

@article{Morbidelli2000,
  author = {Morbidelli, A. and Chambers, J. and Lunine, J. I. and Petit, J. M. and Robert, F. and Valsecchi, G. B. and Cyr, K. E.},
  title = {Source regions and timescales for the delivery of water to Earth},
  journal = {Meteoritics \& Planetary Science},
  volume = {35},
  number = {6},
  pages = {1309--1320},
  year = {2000},
  doi = {10.1111/j.1945-5100.2000.tb01518.x}
}

@article{Lowe2025,
  author = {Drabon, Nadja and Lowe, Donald R. and Byerly, Gary R. and Kyte, Frank T. and others},
  title = {Effect of a giant meteorite impact on Paleoarchean surface environments and life},
  journal = {Proceedings of the National Academy of Sciences},
  volume = {121},
  number = {44},
  pages = {e2408721121},
  year = {2024},
  doi = {10.1073/pnas.2408721121}
}

@article{Schulte2010,
  author = {Schulte, P. and Alegret, L. and Arenillas, I. and and others},
  title = {The Chicxulub asteroid impact and mass extinction at the Cretaceous-Paleogene boundary},
  journal = {Science},
  volume = {327},
  number = {5970},
  pages = {1214--1218},
  year = {2010},
  doi = {10.1126/science.1177265}
}

@article{Boslough2008,
  author = {Boslough, M. B. E. and Crawford, D. A.},
  title = {Low-altitude airbursts and the impact threat},
  journal = {International Journal of Impact Engineering},
  volume = {35},
  number = {12},
  pages = {1441--1448},
  year = {2008},
  doi = {10.1016/j.ijimpeng.2008.07.053}
}

@article{Tancredi2009,
  author = {Tancredi, G. and Hern{\'a}ndez, W. and and others},
  title = {The Carancas meteorite impact},
  journal = {Meteoritics \& Planetary Science},
  volume = {44},
  number = {2},
  pages = {215--233},
  year = {2009},
  doi = {10.1111/j.1945-5100.2009.tb00730.x}
}

@article{Brown2013,
  author = {Brown, P. G. and Assink, J. D. and Astiz, L. and others},
  title = {A 500-kiloton airburst over Chelyabinsk and an enhanced hazard from small impactors},
  journal = {Nature},
  volume = {503},
  number = {7475},
  pages = {238--241},
  year = {2013},
  doi = {10.1038/nature12741}
}

@article{Popova2013,
  author = {Popova, O. P. and Jenniskens, P. and Emel'yanenko, V. and others},
  title = {Chelyabinsk Airburst analysis},
  journal = {Science},
  volume = {342},
  number = {6162},
  pages = {1069--1073},
  year = {2013},
  doi = {10.1126/science.1242642}
}

@techreport{b612annualreport2020,
  author      = {{B612 Foundation}},
  title       = {Asteroid Institute Annual Progress Report 2020},
  institution = {B612 Foundation},
  year        = {2020},
  type        = {Annual Progress Report},
  address     = {Mill Valley, CA, USA},
  url         = {https://b612foundation.org/wp-content/uploads/2020/11/B612AnnualReport2020ReducedSPREADS.pdf}
}

@article{Mainzer2020,
  author = {Mainzer, A. and Bauer, J. and Cutri, R. M. and Grav, T. and Masiero, J. and Beck, R. and others},
  title = {NEOWISE observations of near-Earth objects},
  journal = {The Astronomical Journal},
  volume = {159},
  number = {6},
  pages = {273},
  year = {2020},
  doi = {10.3847/1538-3881/ab89b3}
}

@inproceedings{Hadjara2026SPIE,
  author = {Hadjara, M. and Xu, T. and Lagarde, S. and Ji, H. and Dong, Q. and Diaz, M. and Meftah, M.and Foing, B. and Zhu, P. and Chen, X.},
  title = {From Early Concepts to Lunar Hypertelescopes: Roadmap for Next-Generation Optical Astronomy on the Moon},
  booktitle = {Space Telescopes and Instrumentation 2026: Optical, Infrared, and Millimeter Wave},
  year = {2026},
  editor = {Coyle, Laura E. and Matsuura, Shuji and Perrin, Marshall D.},
  series = {Proceedings of SPIE},
  volume = {\#, (In Press)},
  eid = {14145200},
  pages = {14145200},
  month = jul
}

@article{Harris2015,
  author = {Harris, A. W. and D'Abramo, G.},
  title = {The population of near-Earth asteroids},
  journal = {Icarus},
  volume = {257},
  pages = {302--312},
  year = {2015},
  doi = {10.1016/j.icarus.2015.05.010}
}

@techreport{MPC2026,
  author = {International Astronomical Union Minor Planet Center},
  title = {Minor Planet Center Database Statistics},
  institution = {Harvard-Smithsonian Center for Astrophysics},
  year = {2026},
  url = {https://minorplanetcenter.net}
}

@article{Jedicke2016,
  author = {Jedicke, R. and Granvik, M. and Micheli, M. and others},
  title = {The next decade of Solar System discovery with large synoptic survey telescopes},
  journal = {Icarus},
  volume = {266},
  pages = {173--188},
  year = {2016}
}

@article{Foing2023,
  author  = {Foing, Bernard H. and Lillo, A. and Evellin, P. and Crotts, A. and others},
  title   = {Scientific perspectives on lunar exploration in Europe},
  journal = {Planetary and Space Science},
  year    = {2023},
  volume  = {232},
  pages   = {105701},
  doi     = {10.1016/j.pss.2023.105701}
}
\bibliographystyle{spiebib} 

\end{document}